\documentclass[aps,prl,a4paper,twocolumn,english,superscriptaddress,longbibliography,reprint]{revtex4-1}

\usepackage[mathlines]{lineno}
\usepackage{amsthm}
\usepackage{amsmath}
\usepackage{amssymb}
\usepackage{graphicx}
\usepackage{bm}
\usepackage{color}
\usepackage[dvipsnames]{xcolor}
\usepackage{enumitem}
\usepackage{mathrsfs}
\usepackage{verbatim}
\usepackage{bbold}
\usepackage{braket}

\usepackage{hyperref}

\usepackage{tikz}

\usepackage{natbib}
\usepackage{graphicx}

\newcommand{\new}[1]{#1}
\newcommand{\neww}[1]{#1}

\newcommand{\tr}{\text{Tr}}
\newcommand{\ave}[1]{\langle #1 \rangle}

\begin{document}
\title{From observations to complexity of quantum states via unsupervised learning}
\author{Markus Schmitt}
\email{markus.schmitt@uni-koeln.de}
\affiliation{Department of Physics, University of California, Berkeley, CA 94720, USA}
\affiliation{Institut f\"ur Theoretische Physik, Universit\"at zu K\"oln, 50937 K\"oln, Germany}
\author{Zala Lenar\v{c}i\v{c}}
\affiliation{Department of Physics, University of California, Berkeley, CA 94720, USA}
\affiliation{Jo\v{z}ef Stefan Institute, 1000 Ljubljana, Slovenia}
\date{\today}

\begin{abstract}
The vast complexity is a daunting property of generic quantum states that poses a significant challenge for theoretical treatment, especially in non-equilibrium setups. Therefore, it is vital to recognize states which are locally less complex and thus describable with (classical) effective theories. We use unsupervised learning with autoencoder neural networks to detect the local complexity of time-evolved states by determining the minimal number of parameters needed to reproduce local observations. The latter can be used as a probe of thermalization, to assign the local complexity of density matrices in open setups and for the reconstruction of underlying Hamiltonian operators. Our approach is an ideal diagnostics tool for data obtained from (noisy) quantum simulators because it requires only practically accessible local observations.
\end{abstract}

\maketitle

Finding a suited notion of the complexity of quantum many-body states is a key aspect of various lines of research on correlated quantum systems. Entanglement entropy has emerged as a highly relevant information-theoretic quantity in the case of pure states, that reveals universal features in exotic states of matter \cite{vidal2003,calabrese2004,kitaev2006,fradkin2006,pollmann2010,szasz2020}, in quantum many-body dynamics far from equilibrium \cite{calabrese2005,bardarson2012}, and it indicates the feasibility of classical computer simulations \cite{Scholwoeck2010,Orus2014,Paeckel2019}. Recently, it has also been measured experimentally using cold atoms and trapped ions \cite{islam2015,kokail2020}.
Generalizations of entanglement entropy applicable to mixed states have been proposed \cite{plbnio2007} and entanglement witnesses can detect entanglement \cite{terhal2002}, e.g., in experiments where entanglement entropy is inaccessible.
\new{An alternative perspective is that of quantum circuit complexity, which corresponds to the minimal size of a circuit to prepare the state of interest from a fiducial state \cite{yao1993,dowling08,haferkamp21}.}

\begin{figure}[t]
    \centering
    \begin{tikzpicture}
        \draw[fill=gray!30!white, draw=none] (0,0) rectangle (4,3);
        \draw[fill=orange!40!white, draw=none] (4,0) rectangle (6,3);
        \draw[fill=red!45!lightgray, draw=none] (6,0) rectangle (7.5,3);
        \shade[left color=gray!30!white, right color=orange!40!white] (3.5,0) rectangle (4.5,3);
        \shade[left color=orange!40!white, right color=red!45!lightgray] (5.5,0) rectangle (6.5,3);
        \draw[->, thick] (0,0) -- (7,0);
        \draw[->, thick] (0,0) -- (0,3);
        \node at (3.5,-0.4) {time};
        \node[rotate=90] at (-0.3,1.5) {local complexity};
        \draw[very thick, black] (0,0) 
                                to[out=35,in=225] (1.5,2.25) 
                                to[out=45] (2.25,2.25)
                                to[out=-45, in=165] (4,0.5)
                                to[out=-15,in=180] (6.75,0.25);
        \node at (3,2.5) {\textcolor{white}{(iii)}};
        \node at (5,2.5) {\textcolor{white}{(ii)}};
        \node at (7,2.5) {\textcolor{white}{(i)}};
    \end{tikzpicture}
    \caption{Schematic picture of the ``information barrier'' of quantum many-body dynamics. The different shaded areas illustrate the logical structure of this paper -- we address (i)  infinite time steady states, (ii) practically attainable states at late but finite times, and (iii) short to intermediate times, where the ``information barrier'' is crossed.}
    \label{fig:dynamics_scheme}
\end{figure}
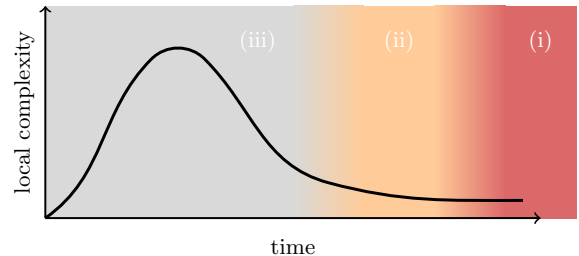

Entanglement \new{and circuit complexity}, however, do not universally align with the physically relevant degree of complexity of many-body states. 
For example, although ergodic many-body systems quickly become highly entangled under unitary non-equilibrium dynamics, the late time dynamics of physically relevant \neww{local observables} is described by hydrodynamic equations of motion of a few quantities \cite{Lux2014,Bohrdt2017,leviatan17}. 
Moreover, their eventual thermalization implies that a few thermodynamic quantities fully characterize local properties, and entanglement entropy turns into thermodynamic entropy of subsystems. Since typical initial states are weakly entangled, it is expected that in terms of \neww{``local complexity'' associated with local observables}, the non-equilibrium dynamics typically passes an ``information barrier'' at intermediate times. 
Such a ``barrier'' has been identified by considering the operator entanglement entropy evolution in closed and open quantum systems \cite{dubail17,wang19,noh20,rakovszky2020,reid21} \new{and the envisioned time-dependence of local quantum state complexity is depicted schematically in Fig.~\ref{fig:dynamics_scheme}}.

This Letter introduces a novel approach to analyze the \neww{local} complexity of quantum many-body states based on recently developed machine learning techniques. At the core, we employ deep autoencoder networks \cite{hinton2006}, which we train in an unsupervised fashion with observable expectation values, in order to obtain a dimensional reduction enabling us to assess the effective complexity of the quantum state of interest. 
We address different regimes of non-equilibrium dynamics of open and closed quantum systems, \new{which are indicated in Fig.~\ref{fig:dynamics_scheme} by the different background shadings}. 
\new{The paper is structured such that we progress from (i) the infinite time limit over (ii) practically attainable steady states at late finite times to (iii) the ``information barrier'' itself at intermediate and short times. Following this order, we can first test our approach in a controlled setting by assigning the local complexity of thermodynamic ensembles; these are idealized local descriptions of steady states of chaotic or integrable models \cite{rigol2007,vidmar16,essler16}. Then, we investigate the dynamical approach of a weakly dissipative system to such states in terms of our complexity measure. For nearly integrable systems, the absence of thermalization despite weak integrability breaking sources is a non-trivial effect \cite{lange17,lange18,reiter19,lenarcic18,lenarcic19}, which can be detected with the unsupervised learning approach.
Finally, we reveal an ``information barrier'' at intermediate times by analyzing the time-dependent local complexity in random unitary dynamics, which has in many recent works been proven to be a useful model to analyze typical features of non-equilibrium dynamics \cite{brown2010,nahum2017,nahum2018,vonKeyserlingk2018,rakovszky2018,khemani2018,chan2018,sunderhauf2018,friedman2019,chan2019,bertini2020,reid21}.}
Besides the assertion of complexity, we demonstrate how further information about effective descriptions can be extracted, e.g., to reconstruct Hamiltonians for thermal states.

Being based on local observations, this approach is suited to exploit highly resolved observations of many-body quantum systems that are possible in modern quantum simulators \cite{leibfried03,Bloch08,islam2015,gross17,preskill18,Ebadi2020,Scholl2020,blais20}.

\begin{figure}[t!]
\includegraphics[width=\columnwidth]{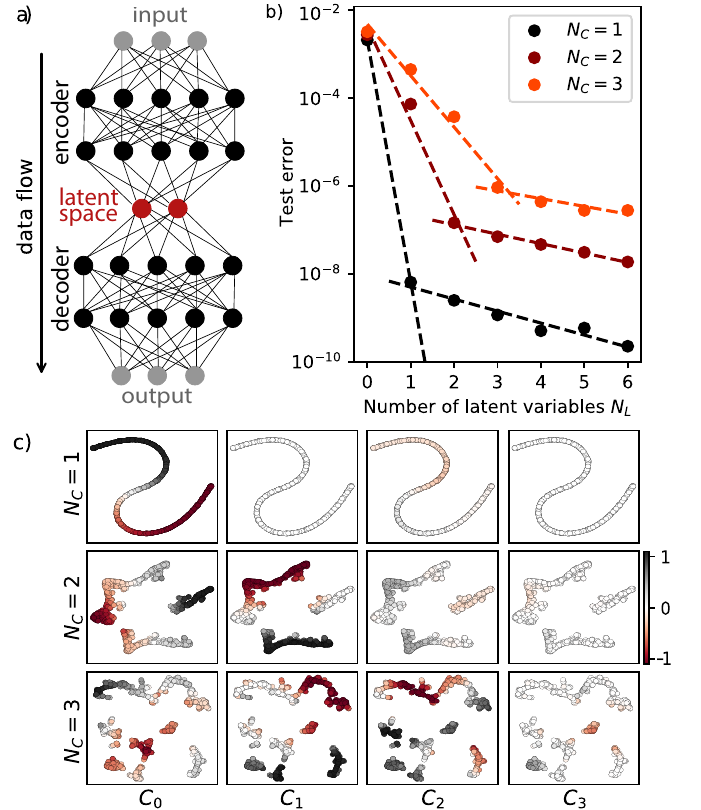}
\caption{a) Schematic depiction of the neural autoencoder network. 
b) Test error as function of the number of latent variables obtained by training on the transverse-field Ising model data from (generalized) Gibbs ensembles with varying number of charges $N_C$. Dashed lines are linear fits included as a guide for the eye.
c) t-SNE of the latent representation of (G)GEs  with different number of charges $N_C$ and $N_L=4$. Color coding in each column is according to expectation values $\ave{\hat C_i}/N$ of charges, denoted in the bottom.
Parameters: $N=12, h_x=0.6J, h_z=0$.
}
\label{fig:fig1}
\vspace{-0.5cm}
\end{figure}

\paragraph{Autoencoder.} 
The autoencoder \cite{hinton2006} is an artificial neural network consisting of multiple layers. The input and output layers have the same dimension, and at least one of the additional layers between the two constitutes a ``bottleneck'' with a considerably smaller number of $N_L$ neurons, spanning the $N_L$ dimensional latent space. The part prior to the bottleneck is called \textit{encoder}, the subsequent part is the \textit{decoder}; see Fig.~\ref{fig:fig1}a. 
Deep learning with autoencoder neural networks has previously been explored as a tool to investigate a variety of physical problems. Examples are unsupervised discovery of physical concepts \cite{iten2020,kottmann2020}, the identification of entangled states \cite{sa2020}, encoding of quantum many-body states \cite{luchnikov2019}, or the exploration of relations to non-equilibrium statistical mechanics \cite{zhong2020}.

Here, the objective of an autoencoder is to reconstruct the input data $x$ given by local observations, despite the intermediate compression in the bottleneck. For this purpose, the reconstruction loss
\begin{equation}
    \mathcal L_{\mathcal D_T}(\boldsymbol\theta)=
    \frac{1}{|\mathcal D_T|}\sum_{x\in\mathcal D_T}
    \big(f_{\boldsymbol\theta}(x)-x\big)^2
    \label{eq:reconstruction_loss}
\end{equation}
over a training data set $\mathcal D_T$ is minimized by optimizing the variational parameters $\boldsymbol\theta$ of the neural network $f_{\boldsymbol\theta}$. Thereby, the encoder learns to map the input data to a suited low-dimensional \textit{latent representation}, which holds the necessary information for the decoder to recover the original input $x$. Achieving a small reconstruction loss depends on the existence of such a low-dimensional representation, and too narrow bottlenecks lead to a larger loss $\mathcal L_{\mathcal D_T}$. Therefore, the effective dimensionality of a data set can be analyzed by varying the bottleneck width and comparing the corresponding reconstruction errors. This analysis should be conducted with validation data that was not used for training to exclude overfitting; we will call the reconstruction error \eqref{eq:reconstruction_loss} evaluated on the test data, the \textit{test error}. Details about the network architecture and optimization are included in the Supplementary Material (SM) \cite{SM}.

\paragraph{Training data.}
In the exemplary applications below we consider quantum spin-$\frac12$ chains, with composite Hilbert spaces $\mathcal H=\bigotimes_l h_l$, where $\text{dim}(h_l)=2$ and $l$ is the lattice site index.
As input for the encoder, we use data sets, where each element consists of the expectation values of all possible operator strings up to a fixed compact support $\mathcal S$, i.e., $O_\rho(\boldsymbol\alpha)=\tr[\hat\rho\hat\sigma_1^{\alpha_1}\ldots\hat\sigma_{|\mathcal S|}^{\alpha_{|\mathcal S|}}]$ with density matrix $\hat\rho$,  $\boldsymbol\alpha=(\alpha_1,\ldots,\alpha_{|\mathcal S|})\in\{0,x,y,z\}^{|\mathcal S|}$ and $\hat\sigma_l^{0,x,y,z}$ denoting the identity and the Pauli matrices, respectively. Measuring these observables amounts to a full tomography of the reduced density matrix of the subsystem $\mathcal S$. While the corresponding cost grows exponentially with subsystem size $|\mathcal S|$, we show in our examples that already moderate supports $|\mathcal S|=3,4$ are informative of the local complexity; for larger supports it might be useful to employ the recently proposed classical shadow techniques \cite{huang20,kokail2020,elben20}.

In our analysis, we consider families of quantum states $\hat\rho_{\boldsymbol\lambda}$ parametrized by a potentially high-dimensional parameter $\boldsymbol\lambda$. 
Individual training data elements are given by observations $\{O_{\rho_{\boldsymbol\lambda}}(\boldsymbol\alpha)|\boldsymbol\alpha\in\{0,x,y,z\}^{|\mathcal S|}\}$ for sampled parameters $\boldsymbol\lambda$.
By optimizing the reconstruction objective \eqref{eq:reconstruction_loss} with autoencoder networks, we gain insights about the effective \neww{local} complexity of $\hat\rho_{\boldsymbol\lambda}$ that is relevant for local observations. The complexity is quantified through the number of latent variables required for faithful reconstruction according to Eq.~\eqref{eq:reconstruction_loss}.
For this procedure it is crucial to exhaustively sample the space of states $\hat \rho_{\boldsymbol\lambda}$ in order to avoid detecting spurious low-dimensional representations.

\paragraph{Idealized steady states of closed systems and Hamiltonian reconstruction.} 
As a first test bed, we consider (generalized) Gibbs ensembles ((G)GEs) of the spin-1/2 quantum Ising model (QIM)
\begin{equation}
\hat H=\sum_{j}\Big( J \hat\sigma_j^z  \hat\sigma_{j+1}^z + h_x \hat\sigma^x_j + h_z \hat\sigma^z_j\Big)\ ,
\label{eq:ising_hamiltonian}
\end{equation}
relevant for quantum simulators  \cite{kim11,islam11,labuhn16,zhang17,Ebadi2020,Scholl2020}. For $h_z=0$, the QIM \eqref{eq:ising_hamiltonian} is integrable, featuring an extensive set of mutually commuting local charges $\hat C_i$ (see SM \cite{SM}), one of which is the Hamiltonian \cite{grady82}. Conservation laws play a crucial role in the long-time description of excited integrable/chaotic systems, \neww{Fig.~\ref{fig:dynamics_scheme}(i)}, when reduced density matrices become indistinguishable from GGEs/GEs of the form \cite{rigol2007,vidmar16,essler16} 
\begin{align}
    \hat\rho_{\boldsymbol\lambda} = \frac{e^{\sum_i\lambda_i\hat C_i}}{\tr[e^{\sum_i\lambda_i\hat C_i}]}.
    \label{eq:thermal_state}
\end{align}
Although long-time states show volume law entanglement entropy \cite{calabrese2005,alba18}, local observables are (for most practical purposes) determined by a few Lagrange parameters $\lambda_i$. 
Our goal is to detect these simple parametrizations.

To benchmark the utility of our approach to analyze such characteristics of long-time states, \neww{ \neww{Fig.~\ref{fig:dynamics_scheme}(i)}, we do not perform an actual time evolution but instead} generate training data from
GGEs with randomly drawn Lagrange multipliers $\lambda_i\in[-2,2]$ using exact diagonalization techniques for the QIM \eqref{eq:ising_hamiltonian} at $h_x=0.6J$, $h_z=0$ with system size $N=12$. 
Each sampled $\boldsymbol\lambda$, yields a training data element with
expectation values of all Pauli strings up to support $|\mathcal S|=3$, and we separate these samples into training and test sets, see SM \cite{SM}. Fig.~\ref{fig:fig1}b displays the test errors, Eq.~\eqref{eq:reconstruction_loss}, achieved after training for different numbers of charges $N_C$ included in the GGEs. We see that the error drops rapidly with an increasing number of latent variables $N_L$ as long as $N_L<N_C$; for $N_L\geq N_C$ the curves level off in all cases. This behavior of the test error is expected because $N_C$ is the minimal number of independent variables needed for encoding the GGE data. The change of slope becomes more distinct with larger training data sets \cite{SM}.

In Fig.~\ref{fig:fig1}c we employ t-distributed stochastic neighbor embedding (t-SNE) \cite{van08} to visualize what the autoencoder learned to encode in its latent space. The t-SNE is a dimensional reduction technique, where the proximity of data points in the resulting low-dimensional representation is determined based on their Euclidian distance in the original space. The position of each point corresponds to the latent representation of one set of observations. The color code indicates the corresponding expectation value of the different charges. For $N_C=1$, we find that also for a higher-dimensional latent space, $N_L>N_C$, 
the latent representation of observations lies on a one-dimensional manifold with energy density monotonously changing along it.
When including more charges, we see that the dimensionality of the latent representation grows. For $N_C>1$, the locations of extremal regions of the color code reveal that each charge can be associated with a different direction in the latent space. This indicates that the unsupervised learning procedure yielded an encoding directly related to the physical charges.
In the SM \cite{SM}, we show that the learned representation is connected to the charges through an invertible map, which is sufficient for an unambiguous encoding of the data.

Having demonstrated the possibility to identify and analyze ideal (G)GE states, we next turn to genuine late-time states, \neww{Fig.~\ref{fig:dynamics_scheme}(ii)}. 
\neww{
To add some additional structure to these states, we perform time evolution with respect to open quantum systems.
As we show in the following, the local complexity of late-time states in closed and weakly open systems share many properties.}

\paragraph{Steady states of open systems.} Here we characterize steady states of open systems described by the Liouville equation
\begin{equation}\label{eq::NESS}
\dot{\hat \rho} = i[\hat \rho,\hat H] + \epsilon\sum_{j,\gamma}\big(\hat L^{(\gamma)}_j \hat\rho \hat L_j^{(\gamma)\dagger} - \frac{1}{2}\{\hat L_j^{(\gamma)\dagger} \hat L_j^{(\gamma)}, \hat\rho\}\big)    
\end{equation}
with couplings to non-equilibrium baths represented by Lindblad operators $\hat L_j^{(\gamma)}$. The strength of the Markovian part is parametrized by $\epsilon$. 
\neww{In direct relation with the previous section},
at a weak coupling to dissipators, $\epsilon \ll 1$, steady states for a chaotic (integrable) $\hat H$ can be approximated with (G)GEs, Eq.~\eqref{eq:thermal_state}, which are corrected by a small $\delta\hat \rho(\epsilon)$, $\hat \rho \propto \hat \rho_{\boldsymbol{\lambda}} + \delta\hat\rho(\epsilon)$.
For an integrable $\hat H$, the absence of thermalization despite weak integrability breaking sources is a non-trivial effect observed in weakly open systems \cite{lange17,lange18,reiter19,lenarcic18,lenarcic19}.

In this context, the unsupervised learning approach enables us to address several natural questions.
For the integrable $\hat H$, we will utilize our approach to detect how many conservation laws must be considered to reproduce local observables -- a question often raised in quench protocols in closed setups. For the chaotic $\hat H$, we will examine the effect of noise as of relevance for quantum simulators that are never completely isolated.

\begin{figure}[t!]
\includegraphics[width=\columnwidth]{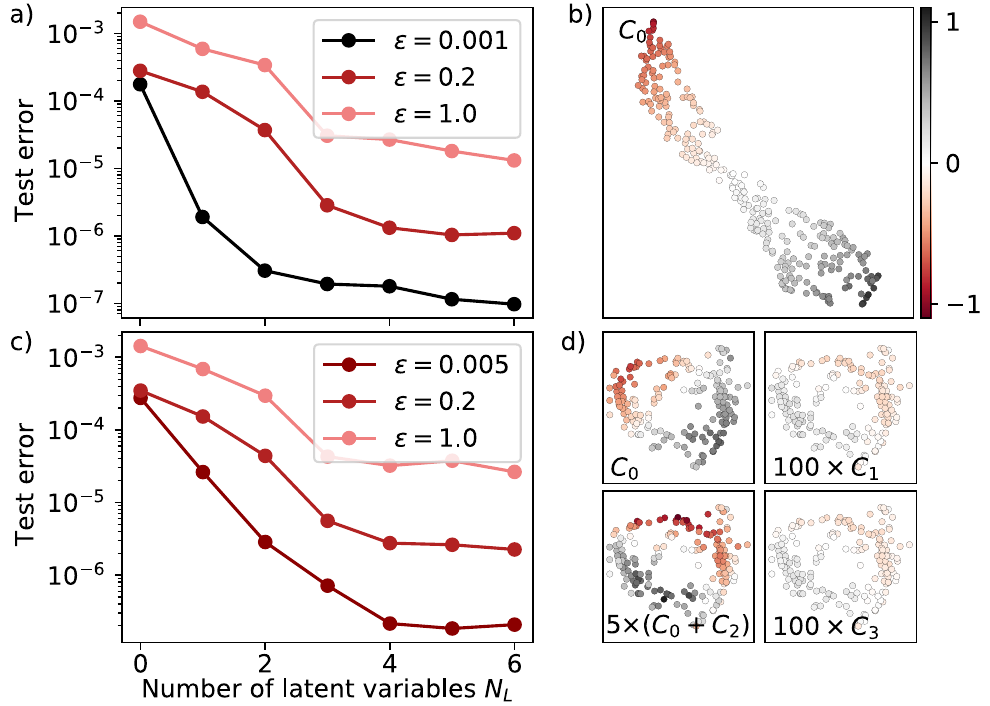}
\caption{
Results obtained by training on data from steady states for setups with different strength of openness $\epsilon$, Eq.~\eqref{eq::NESS}, for chaotic (top) and integrable (bottom) $\hat H$.
a,c) Test error as function of the number of latent variables.  
b) t-SNE of the latent representation for chaotic $\hat H$ and $\epsilon=0.001$, $N_L=5$. Color coding is with respect to energy density. 
d) t-SNE of the latent representation for integrable $\hat H$ and $\epsilon=0.005$, $N_L=5$. Color coding is with respect to  (rescaled) densities of charges written in the inset. The gradient in the coloring with respect to $(5/N)\ave{\hat C_0 + \hat C_2}$ is roughly perpendicular to the gradient of $\ave{\hat C_0}/N$. For our choice of Lindblads, inversion non-symmetric $\hat C_{1,3}$ show much smaller expectation values, i.e., not comparable to $\ave{\hat C_0}/N$ even upon rescaling with 100 and are thus not so crucial for an approximate description. 
Parameters: $N=40$, $h_x=1.152J$, $h_z=0.974J$ (top) and $h_x=1.352J$ (bottom).
}
\label{fig:fig2}
\end{figure}

The Hamiltonian $\hat H$ is again the QIM, Eq.~\eqref{eq:ising_hamiltonian}, while the Lindblad dissipators \neww{for different data elements} are  randomly rotated single and two-site operators, see SM \cite{SM}. 
The exact form of $\hat L_j^{(\gamma)}$ is not important, but sufficient randomness in either the form or the relative dissipation rates is needed to have a diverse enough training set.
Fig.~\ref{fig:fig2} shows results for steady states obtained using time-evolving block decimation technique \cite{zwolak04,verstraete04} for vectorized density matrices on systems of size $N=40$, using bond dimension $\chi=100$. For $|\mathcal S|=4$, we plot the test error as a function of the number of latent variables for different strengths of coupling to Markovian baths $\epsilon$ and for Hamiltonians that are either integrable or non-integrable.

At a weak coupling to baths $\epsilon=0.001$ and chaotic $\hat H$ the test error levels off at $N_L\approx2$, Fig.~\ref{fig:fig2}a. In the SM \cite{SM}, we show that the single latent variable, which already reaches high accuracy, again corresponds to the energy, confirming that Gibbs ensembles approximate density matrices. 
Further latent variables capture information about $\delta\hat\rho(\epsilon)$ corrections. 
We find that observables strongly and monotonously varying along the perpendicular direction of the tSNE representation are related to the features of the baths, \neww{i.e., the correlations that baths promote,} 
see SM \cite{SM}.
This could be used for the detection of noise type, as relevant for quantum error correction.

For integrable $\hat H$, Fig.~\ref{fig:fig2}c, more than one latent variable is needed even for an approximate description. The t-SNE in Fig.~\ref{fig:fig2}d reveals that the first two latent variables are related to linear combinations of the two most local inversion symmetric charges $\hat C_0=\hat H$ and $\hat C_2$, confirming that GGEs approximately describe steady states \cite{lange17,lange18,reiter19}. Furthermore, our approach exposes that of macroscopically many conservation laws, only two are particularly relevant for an approximate description of all local observables with support up to $|\mathcal{S}|=4$. 
Further latent variables are for corrections $\delta\hat\rho(\epsilon)$. At stronger $\epsilon=0.2,1.0$, a larger number of latent variables is necessary. Hence, for our choice of Lindblad operators, there is no simple emergent description. 

\begin{figure}
\includegraphics[width=\columnwidth]{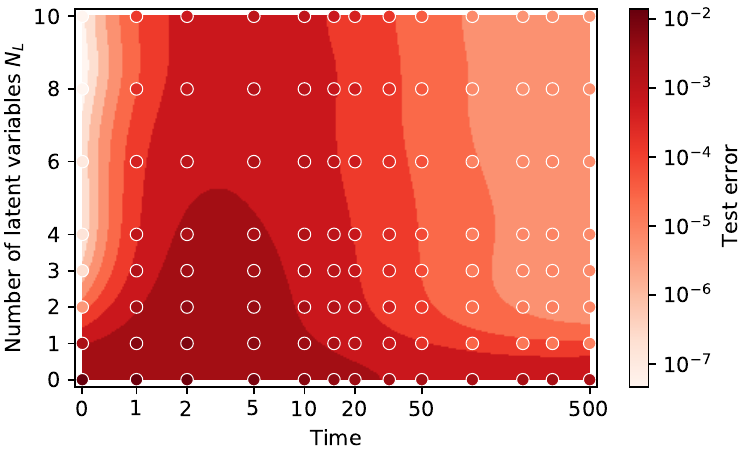}
\caption{Complexity during random unitary evolution of $N=20$ spins captured by the autoencoder network. Starting from a simple product state, an intermediate regime of high complexity is traversed before an effective one-dimensional description emerges at late times due to the presence of a conserved charge.
For times $t>1$ the scale is logarithmic. Data points (colored dots) are augmented by discretized heatmap based on a bivariate B-spline interpolation.}
\label{fig:rnd_unitary}
\end{figure}

\paragraph{Random unitary evolution and information bottleneck.} 

\neww{Finally, we study the local complexity at intermediate times, Fig.~\ref{fig:dynamics_scheme}(iii),  by analyzing states evolved by random unitary dynamics.}
Random unitary circuits have recently received substantial interest because they allow the analysis of typical features of non-equilibrium dynamics \cite{brown2010,nahum2017,nahum2018,vonKeyserlingk2018,rakovszky2018,khemani2018,chan2018,sunderhauf2018,friedman2019,chan2019,bertini2020,reid21}. We will now show that in combination with our unsupervised learning of \neww{local} complexity, they provide suited grounds to explore the information barrier, appearing at intermediate times in the time evolution from simple (product) states.
\neww{To detect the information barrier, a detour away from Hamiltonian dynamics is necessary, in order to add a random component to the time evolution.} From the perspective of the autoencoder, states that are obtained through deterministic dynamics starting from the few-parameter family of product states have a constant complexity: the fixed unitary map can be absorbed in the encoder/decoder of the network such that a few-parameter representation is possible at all times. \neww{On the other hand,} the effective reduction of complexity at late times could also be studied under Hamiltonian dynamics when considering a higher-dimensional manifold of initial states.

Our model of random unitary evolution is based on alternating application of two-site gates on even and odd links of the spin chain. The unitary gates vary randomly from step to step, but they are homogeneous across the chain to preserve translational symmetry. Moreover, we choose the unitaries to be $U(1)$-symmetric \neww{to impose magnetization as a single conserved quantity. Building on the results of the previous sections, we anticipate that a single Lagrange multiplier associated with the magnetization conservation will locally characterize states at late times.}
The initial states are random translationally invariant product states. At each time step, we measure all operators with support $|\mathcal{S}|=3$ and train the autoencoder with different $N_L$-dimensional bottlenecks. See SM \cite{SM} for details.

Fig.~\ref{fig:rnd_unitary} displays the test errors 
that we find as a function of time and number of latent variables. The full description of the initial product state by two Bloch angles is reflected in very small test errors when using two or more latent variables. 
At late times a single latent variable is sufficient for faithful reconstruction, consistent with convergence to a reduced density matrix corresponding to an ensemble fully characterized by a single Lagrange multiplier. 
At intermediate times, however, 
the reconstruction error becomes large even for rather wide bottlenecks because the generic dynamics modelled by random unitary circuits generates complex quantum states, for which no effective low-dimensional parametrization exists. 
In that sense, our unsupervised learning approach probes the information barrier of quantum many-body dynamics. \neww{This concept of local complexity should, on the other hand, not be mistaken for the circuit complexity, which generically grows as a function of the random circuit depth \cite{haferkamp21}.}

\paragraph{Hamiltonian reconstruction.}
\new{When our approach reveals that data has a (near) one-dimensional latent representation (like top Fig.~\ref{fig:fig1}c or Fig.~\ref{fig:fig2}b), presumably corresponding to thermal states $\rho$, this connection can be exploited to reconstruct $\hat H$, as we outline now for the translationally invariant case. We first single out operators $\hat O(\boldsymbol{\alpha})\equiv \hat O_{j=1}(\boldsymbol{\alpha})=\hat\sigma_1^{\alpha_1}\ldots\hat\sigma_{|\mathcal S|}^{\alpha_{|\mathcal S|}}$ for which we measure a large average gradient of $O_\rho(\boldsymbol{\alpha})=\tr[\hat\rho \, \hat O(\boldsymbol{\alpha})]$ along the effective one-dimensional latent manifold, spanned by different temperatures. These are candidate Hamiltonian terms $\hat H=\sum_{j,\boldsymbol{\alpha}} a_{\boldsymbol{\alpha}}\hat O_j(\boldsymbol{\alpha})$. Using Newton's method or similar, we find coefficients $a_{\boldsymbol{\alpha}}$ up to a scaling factor from the conditions $\tr[\hat O(\boldsymbol{\alpha}) \, e^{\sum_{j,\boldsymbol{\alpha}} a_{\boldsymbol{\alpha}} \hat O_j(\boldsymbol{\alpha})}]/\tr[e^{\sum_{j,\boldsymbol{\alpha}} a_{\boldsymbol{\alpha}} \hat O_j(\boldsymbol{\alpha})}] -  O_\rho(\boldsymbol{\alpha})=0$. Testing this approach with the data presented in Fig.~\ref{fig:fig1}, $h_x/J$ is reconstructed within the accuracy of the Newton's method. Using the data in Fig.~\ref{fig:fig2}b for $\epsilon=0.001$, the relative strength of fields is obtained within 5\% and 1\% for $h_x/J$ and $h_z/J$, respectively. See the SM \cite{SM} for more information on the algorithm. In contrast to other methods \cite{che2020,valenti19,ma20,xin19}, our approach is unsupervised and relies only on expectation values.}

\paragraph{Discussion.}
We have introduced an unsupervised learning approach that reveals information about the existence of effective low-dimensional descriptions of many-body quantum states.
Although no domain knowledge is provided during optimization, we showed for some examples how physical information can be extracted from the trained autoencoders. In that sense, our results constitute a step towards interpretable machine learning of physics. 

An alternative to our approach is the analysis of the intrinsic dimension of the data sets \cite{wold87,borg05,balasubramanian02,van08,mcinnes18,levina04,campadelli15,facco17,mendes20,lopez18,turkeshi21,aiman21}. In the SM \cite{SM} we include a corresponding analysis of our data as reference. While the results are consistent, the autoencoder method seems more robust, and it has the advantage to reveal further information beyond the minimal number of independent variables.

Autoencoders could be used, for example, to detect GGEs in weakly open trapped-ions setups \cite{reiter19}.
Natural extensions of the present work could be the investigations of Floquet prethermal plateaus and of thermalization in strongly disordered many-body systems. Our method could complement the recent proposal to use supervised or confusion learning based on observations from cold atom experiments \cite{bohrdt2020}.
In the SM \cite{SM}, we give first evidence that autoencoders can be used for noise-type recognition, relevant for quantum error correction, which opens another exciting direction.
Finally, interesting questions to explore are the possibility of learning with incomplete or imperfect observations. In that regard, refinements of the deep learning model might be required, e.g., the utilization of variational autoencoders for resilience against noise \cite{diederik2019}.

\begin{acknowledgments} 
We thank M. Bukov for valuable comments on the manuscript and M. Dalmonte for constructive remarks on the intrinsic dimension calculation. Z.L. acknowledges also discussions with O. Alberton.
Our TEBD code was written in Julia \cite{bezanson17}, relying on the TensorOperations.jl. For data from exact diagonalization we used the Quspin library \cite{quspin1,quspin2}. The used training data, as well as code for training and analysis are available online \cite{github}.
M.S. was supported through the Leopoldina Fellowship Programme of the German National Academy of Sciences Leopoldina (LPDS 2018-07) with additional support from the Simons Foundation.
Z.L. was financed by Gordon and Betty Moore Foundation’s EPIC initiative, Grant No. GBMF4545 and J1-2463 of the Slovenian Research Agency.
\end{acknowledgments}

\bibliographystyle{apsrev4-1}
\bibliography{references}

\end{document}

\title{\textit{Supplemental material to}\\\vspace{5pt}From observations to complexity of quantum states via unsupervised learning}
\author{Markus Schmitt}
\affiliation{Department of Physics, University of California, Berkeley, CA 94720, USA}
\affiliation{Institut für Theoretische Physik, Universität zu Köln, 50937 Köln, Germany}
\author{Zala Lenar\v{c}i\v{c}}
\affiliation{Department of Physics, University of California, Berkeley, CA 94720, USA}
\affiliation{Jo\v{z}ef Stefan Institute, 1000 Ljubljana, Slovenia}
\date{\today}

\maketitle
\section{Conservation laws of transverse field Ising model}
The transverse field Ising model, a special case of the quantum Ising model (QIM) used in the main text, 
\begin{equation}\label{eq::SM-TFIM}
\hat H=\sum_{j} J \hat\sigma_j^z  \hat\sigma_{j+1}^z + h_x \hat\sigma^x_j
\end{equation}
is an example of non-interacting integrable model. If fermionized, the fermionic mode occupations are its non-local conservation laws. One can, alternatively, construct local conservation laws of the following form \cite{grady82}
\begin{align}\label{eq:sm:charges}
\hat C_0&=\hat H \notag \\
\hat C_2&=\sum_j 
J\hat S^{zz}_{_{j,j+2}}
- h_x \hat\sigma^y_{j}\hat\sigma^y_{j+1} 
- h_x \hat\sigma^z_{j} \hat\sigma^z_{j+1}
- J \hat\sigma^x_{j} \notag \\
\hat C_{2k>2}&=  \sum_j 
J \hat S^{zz}_{_{j,j+k+1}} 
- h_x \hat S^{yy}_{_{j,j+k}}
- h_x \hat S^{zz}_{_{j,j+k}}
+J \hat S^{yy}_{_{j,j+k-1}} \notag \\
\hat C_{2k-1}&= J \sum_j \hat S^{yz}_{_{j,j+k}} - \hat S^{zy}_{_{j,j+k}},
\end{align}
where $\hat S^{\alpha\beta}_{i,j}=\hat\sigma^\alpha_i \hat\sigma^x_{i+1} \dots\hat\sigma^x_{j-1} \hat\sigma^\beta_{j}$. Such a formulation of conservation laws is useful, because there typically exists a hierarchy of importance in the description of equilibration in integrable systems: expectation values of local observables are typically reasonably well captured by truncated generalized Gibbs ensembles that include only the most local conservation laws \cite{essler16}.

Also in our study of QIM in the presence of weak openness and driving, t-SNE visualization of what was learned in the latent space pointed out that the two principal directions were spanned by a linear combination of $\hat C_0=\hat H$ and $\hat C_2$, i.e., the expectation values of all operators with support up to 4 were given mainly by $\ave{\hat H}$ and $\ave{\hat C_2}$, while less local conservation laws were less crucial. While our choice of Lindblad driving, Eq.~\eqref{eq::Lind1}, was not parity symmetric, and thus $\ave{\hat C_{2k-1}}\neq 0$ and $\lambda_{2k-1}\neq 0$, the importance of $\hat C_1$ and $\hat C_3$ was clearly smaller. Coupling to baths that are by symmetry geared towards generation of currents would, on the other hand, promote the importance of $\ave{\hat C_{2k-1}}$ \cite{lange18}.

\section{From Lagrange parameters to latent representation}
An ideal test bed for the discussed methods are data sets of expectation values calculated from generalized Gibbs ensembles (GGEs). These are statistical density matrices of the form
\begin{align}\label{eq:sm:GGE}
\hat\rho_{\boldsymbol{\lambda}}=\frac{1}{Z(\boldsymbol\lambda)}\exp\bigg(\sum_{i=0}^{N_C-1}\lambda_i\hat C_i\bigg),
\end{align}
defined by a set of $N_C$ mutually commuting conserved charges $\hat C_i$, $[\hat C_i,\hat C_j]=0$ and fully characterized by $N_C$ parameters $\lambda_i$, which are called Lagrange multipliers. Lagrange multipliers of common charges have special names, for example the temperature or the chemical potential. Moreover, $Z(\boldsymbol\lambda)=\tr[e^{\sum_i\lambda_i \hat C_i}]$ denotes the partition sum.

If the expectation values of the charges are known, the Lagrange multipliers are implicitly determined by
\begin{align}
    \langle\hat C_i\rangle=\tr\big[\hat C_i\hat\rho_{\boldsymbol\lambda}\big]\equiv F_i(\boldsymbol\lambda) \ .
    \label{eq:implicit_lagrange_multipliers}
\end{align}
In order to obtain the Lagrange multipliers one has to solve this system of non-linear equations, i.e., find $F_i^{-1}$ such that
\begin{align}
    \lambda_i=F_i^{-1}(\boldsymbol C)\ ,
    \label{eq:lagrange_inverse}
\end{align}
where $\boldsymbol C=(\langle\hat C_1\rangle,\ldots,\langle\hat C_{N_C}\rangle)$.

Generally, the charges can be written as linear combination of the local operator strings,
\begin{align}
    \hat C_i=\sum_{\boldsymbol\alpha,j}c_{i\boldsymbol\alpha}\hat O_j(\boldsymbol\alpha)
\end{align}
We can expand the set of equations \eqref{eq:implicit_lagrange_multipliers} to include all operator strings 
$\hat O(\boldsymbol\alpha)
\equiv \hat O_1(\boldsymbol\alpha)
=\hat\sigma_1^{\alpha_1}\ldots\hat\sigma_{|\mathcal S|}^{\alpha_{|\mathcal S|}}$ of a given support $|\mathcal S|$, characterized with $\boldsymbol\alpha=(\alpha_1,\ldots,\alpha_{|\mathcal S|})\in\{0,x,y,z\}^{|\mathcal S|}$, yielding
\begin{align}
    \langle\hat O(\boldsymbol\alpha)\rangle&=\tr\big[\hat O(\boldsymbol\alpha)\hat\rho_{\boldsymbol\lambda}\big]\equiv \tilde F_{\boldsymbol\alpha}(\boldsymbol\lambda)\label{eq:expectation_value_rho}
\end{align}
Therefore, using Eq.~\eqref{eq:lagrange_inverse}, we can write the inverse of Eq.~\eqref{eq:expectation_value_rho} as
\begin{align}
    \lambda_i&=F_i^{-1}\Big(\sum_{\boldsymbol\alpha}c_{i\boldsymbol\alpha}\braket{\hat O(\boldsymbol\alpha)}\Big)\equiv\tilde F_i^{-1}(\boldsymbol O)\ .
\end{align}
Now $\boldsymbol O$ is a vector of all measured operators with different $\boldsymbol\alpha$ and corresponds to the type of data set introduced in the main text. Moreover, the self-consistency equation
\begin{align}
    \langle\hat O_i\rangle=\tilde F_i\big(\tilde F_1^{-1}(\boldsymbol{O}),\ldots,\tilde F_{N_C}^{-1}(\boldsymbol{O})\big)
\end{align}
resembles the structure of the bottleneck autoencoder: $\tilde F_{i}^{-1}$ is the non-linear encoder that reduces the input data to a low-dimensional latent representation $\boldsymbol\lambda$, and $\tilde F_{i}$ is the non-linear decoder that reproduces the input data from the given latent representation. In the main text, we showed that for data from GGEs the unsupervised training of bottleneck autoencoders indeed allows to identify meaningful low-dimensional representations related to the charges and Lagrange multipliers.

\begin{figure}
\includegraphics[width=0.9\columnwidth]{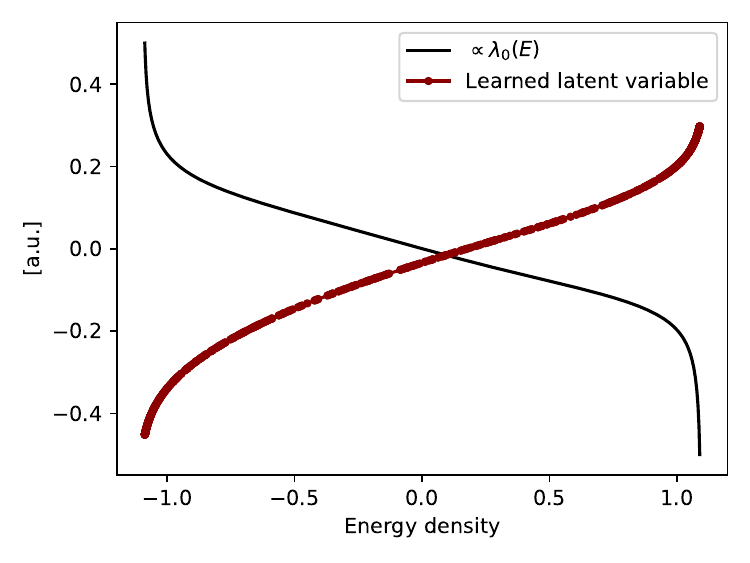}
\caption{The learned latent representation of a network with a bottleneck of width one as function of energy density compared to the inverse temperature of the corresponding density matrix.
}
\label{fig:latent_vs_energy}
\end{figure}

Here we give some additional information on this relation.
First, we consider the thermal density matrices
\begin{align}
    \hat\rho_{\lambda_0}=\frac{1}{Z(\lambda_0)}\exp\bigg(\lambda_0\hat H\bigg)
    \label{eq:sm:GE}
\end{align}
with $\hat H$ the QIM Hamiltonian \eqref{eq::SM-TFIM} at $h_x/J=0.6$. As shown in the main text, Fig.~1, we compute the expectation values \eqref{eq:expectation_value_rho} using exact diagonalization on $N=12$ sites. For this purpose we exploit translational symmetry to reduce the dimension and work in the zero momentum sector. Considering all possible operators $\hat O(\boldsymbol{\alpha})$ supported on no more than $|\mathcal{S}|=3$ lattice sites, we computed the expectation values for random temperatures, drawn from the uniform distribution on the interval $\lambda_0\in [-2/J,2/J]$.

Using a bottleneck network with a single latent variable, we achieve a test error $\mathcal L_{\text{test}}\approx10^{-8}$. This indicates that a very accurate one-dimensional encoding is found, as we would expect, because the density matrix \eqref{eq:sm:GE} is fully characterized by temperature as a single parameter. In Fig.~\ref{fig:latent_vs_energy} we show for our full data set the value attained by the latent variable as a function of energy. We include also the corresponding inverse temperatures. A naive expectation would be that during training the bottleneck networks learns to encode either the temperature or the energy density in the available latent variable. If temperature is encoded, the decoder would have to learn only the function $\tilde F(\lambda_0)$ as in Eq.~\eqref{eq:expectation_value_rho}. If energy is encoded, the decoder would also need to include the inversion as in Eq.~\eqref{eq:lagrange_inverse}, i.e., approximate $\tilde F\circ F^{-1}$. The result in Fig.~\ref{fig:latent_vs_energy}, however, shows that the latent representation learned does not coincide with either of the two. And, in fact, it is sufficient to learn a latent representation $\boldsymbol l$ that is related to energy (or temperature) by an invertible function $G$, because with $E=G^{-1}(\boldsymbol l)$ the decoder can learn $\tilde F\circ F^{-1}\circ G^{-1}$ in order to map the latent value $\boldsymbol l$ to the vector of observations $\boldsymbol O$. In Fig.~\ref{fig:latent_vs_energy} we see that $\boldsymbol l=G(E)$ is strictly monotonic and, hence, indeed invertible.

\begin{figure}
\includegraphics[width=\columnwidth]{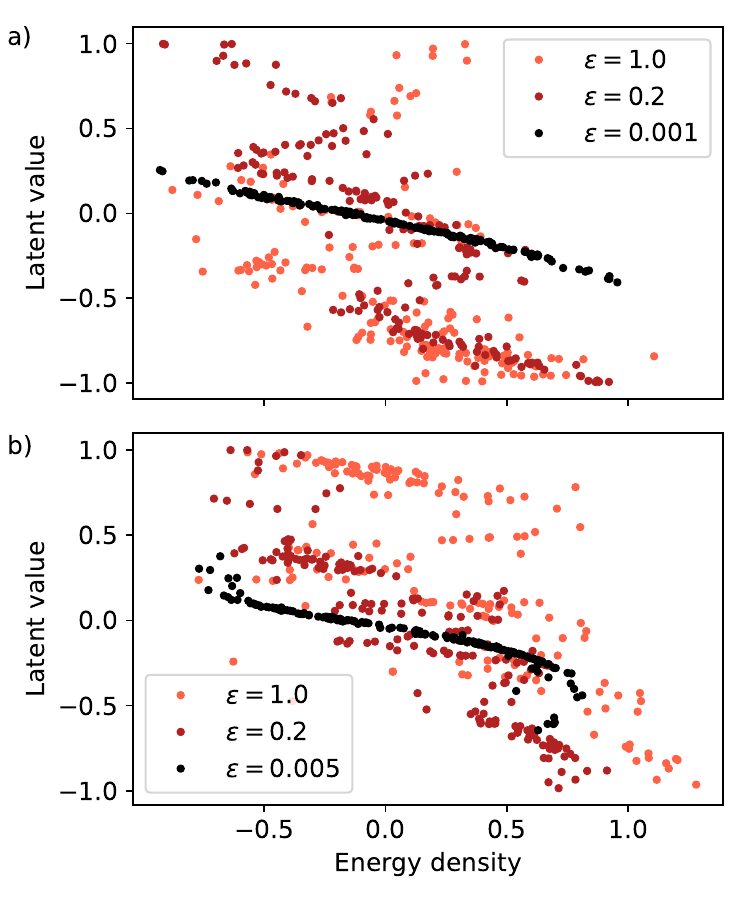}
\caption{
The learned latent representation as a function of energy density, obtained by a network with a single latent variable for different strength of coupling to the environment $\epsilon$ and (a) chaotic or (b) integrable $\hat H$. (a) Only for $\epsilon=0.001$ and chaotic $\hat H$, energy and the first latent variable are related with an invertible function. (b) For $\epsilon=0.005$ and integrable $\hat H$, the first latent variable and energy are strongly correlated, however, non-monotonicity indicates that further latent variables, associated with additional charges, are needed. At stronger couplings to baths $\epsilon=0.2,1.0$ the first latent variable is not related to energy at all.
\label{fig:SM:Lindblad}}
\end{figure}

In an open Markovian setup, such a relation exists in the limit of weak coupling to the environment. In the main text, we consider steady states of combined unitary and Markovian dynamics
\begin{equation}
\dot{\hat\rho} = i[\hat\rho,\hat H] + \epsilon\sum_{j,\gamma}\big(\hat L^{(\gamma)}_j \hat \rho \hat L_j^{(\gamma)\dagger} - \frac{1}{2}\{\hat L_j^{(\gamma)\dagger} \hat L_j^{(\gamma)}, \hat\rho\}\big). 
\end{equation}
The unitary part is given by the QIM, Eq.~\eqref{eq::SM-TFIM}, potentially in an additional integrability breaking $z$-field, $h_z \sum_j \hat\sigma_j^z$. Bulk Lindblad operators $L_j^{(\gamma)}$  are rotated single ($\gamma=1$) and two-site ($\gamma=2$) operators,
\begin{align}\label{eq::Lind1}
\hat L_j^{(1)} &= R_z(\zeta')R_y(\phi')\frac{\hat \sigma^0_{j}+\hat \sigma^z_{j}}{2}R^{-1}_y(\phi')R^{-1}_z(\zeta'),\\
\hat L_j^{(2)} &= \left(R_z(\zeta)R_y(\phi)\hat S^{-}_{j-1}R^{-1}_y(\phi)R^{-1}_z(\zeta)\right) \frac{\hat\sigma^0_{j}-\hat\sigma^z_{j}}{2}. \notag
\end{align}
Here, $R_{z(y)}(\zeta)$ is a rotation for $\zeta$ around the $z(y)$ spin axis and $\zeta, \zeta',\phi,\phi'\in[-\pi,\pi]$ are chosen randomly.

At a weak coupling to the environment, $\epsilon\ll 1$ and a chaotic $\hat H$, the first latent value is still related to the energy $\tr[\hat H \hat\rho]$ by an invertible function, as seen in Fig.~\ref{fig:SM:Lindblad}a, showing results obtained with a single latent variable network. In our setup, the first latent variable turns out to be largely related to energy also for the integrable $\hat H$, Fig.~\ref{fig:SM:Lindblad}b. The fact that further latent variables, related to other charges, are needed, is revealed only by extremal values of energy and latent values, which show that the relation is nonetheless non-invertible in this case.
As we pointed out in the main text, for stronger couplings $\epsilon=0.2,1.0$ to baths, steady states are no longer approximately described by Gibbs ensembles of form \eqref{eq:sm:GE}, thus energy plays no unique role.

\section{Detecting the structure of noise}

\begin{figure}[b!]
\includegraphics[width=\columnwidth]{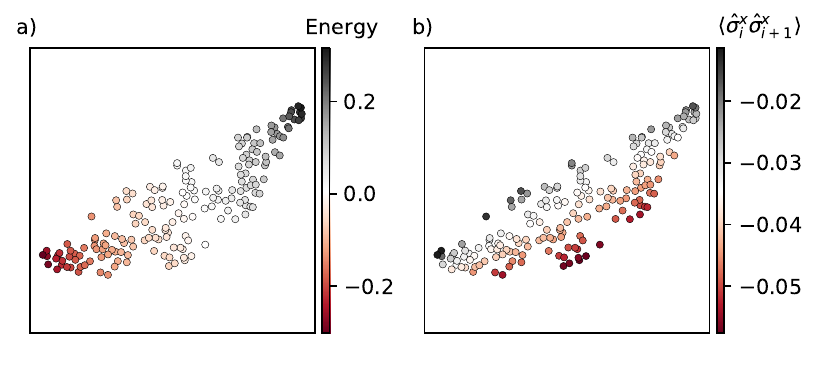}
\caption{
tSNE for the latent representation of steady state expectation values in setup with chaotic QIM and Eq.\eqref{eq:L_x} Lindblad operators. The coloring is with respect to (a) the energy density and (b) correlations $\ave{\hat\sigma_j^x\hat\sigma_{j+1}^x}$. This reveals that the first latent variable is related to energy, while the second one detect the type of correlations that our structured noise is promoting. Parameters: $h_x/J=0.709$, $h_z/J=0.9042$, $\epsilon=0.02$, $N=40$, $\chi=100$, $|\mathcal{S}|=4$.
}
\label{fig:structured_noise}
\end{figure}

In the main text and above we showed that in the latent representation of steady states for weakly open system with chaotic $H$, the first latent variable is related to the energy density (or temperature). This is in agreement with the steady state $\hat\rho(\epsilon) \sim \hat\rho_{\lambda_0} + \delta\hat\rho(\epsilon)$ being approximated by a Gibbs ensemble $\hat\rho_{\lambda_0}$. However, further latent variables are needed to capture corrections $\delta\hat\rho(\epsilon)$. If baths have some structure, i.e., are promoting certain correlations, those correlations can be extracted from the additional latent variables.

To show this we consider chaotic QIM, weakly coupled to structured baths represented by dissipators
\begin{equation}
\cl{D}
=\epsilon 
\big( \gamma_1 \cl{D}^{(1)} + \gamma_2\cl{D}^{(2)} + \gamma_3 \cl{D}^{(3)} + \gamma_4\cl{D}^{(4)} + 	\cl{D}^{(5)}\big)
\end{equation}
with random rates $\gamma_1, \gamma_2,\gamma_3,\gamma_4\in[0,1]$ and of the form given by Lindblad operators
\begin{align}\label{eq:L_x}
\hat L_j^{(1)} &= \hat S_j^{+,x} \hat P^{\downarrow,x}_{j+1} , \quad
\hat L_j^{(2)} = \hat P^{\downarrow,x}_{j} \hat S_{j+1}^{+,x}, \\
\hat L_j^{(3)} &= \hat S_j^{-,x} \hat P^{\uparrow,x}_{j+1},\quad 
\hat L_j^{(4)} = \hat P^{\uparrow,x}_{j} \hat S_{j+1}^{-,x},\notag \\
\hat L_j^{(5)} &= \hat S_j^z \notag
\end{align}
where $\hat S^{\pm,x}_j=\frac{1}{2}(-\hat\sigma^z_j\pm i\, \hat\sigma^y_j)$
and $\hat P_j^{\uparrow/\downarrow,x} = \frac{1}{2}(\hat\sigma^0_j \pm \hat\sigma^x_j )$.
It is clear that while $\hat L_j^{(5)}$ cause featureless dephasing, $\hat L_j^{(1-4)}$ enhance anti-ferromagnetic correlations in spin $x$-direction, $\ave{\sigma_j^x\sigma_{j+1}^x}<0$. 

Given that the latent representation provides compressed information on the steady state, we can use it to detect antiferromagnetic correlations in the data. Fig.~\ref{fig:structured_noise} shows tSNE of the latent representation, colored by (a) the energy density and (b) correlation $\ave{\sigma_j^x\sigma_{j+1}^x}$. Like for the other choice of Lindblad operators, the first latent variable is related to energy, resulting in tSNE being primarily extended along this direction, Fig.~\ref{fig:structured_noise}a. As shown in Fig.~\ref{fig:structured_noise}b, the perpendicular direction is indeed spanned by the antiferromagnetic correlations $\ave{\sigma_j^x\sigma_{j+1}^x}<0$, promoted by the bath. The second latent variable is thus capturing the most crucial information in $\delta\hat\rho(\epsilon)$, namely, the anti-ferromagnetic correlations. 

As we demonstrated above, our approach has the potential do detect the form of structured noise that might be present in a quantum simulator or quantum computer, at each run with a different strength but of the same form. Knowing it could assist quantum error correction algorithms.

\section{Random unitary model}
We consider a model of random unitary dynamics of a spin-1/2 chain with $N=20$ sites that preserves the total magnetization. In particular, we consider random two-site unitaries that are applied alternatingly on even and odd links of the chain. This approach preserves locality of dynamics despite the randomness and it was studied extensively in previous works \cite{nahum2017,nahum2018,vonKeyserlingk2018,rakovszky2018}.
However, our model differs slightly from the conventional Haar random unitaries, because it is constructed with two constraints in mind, namely (i) the ability to work with a small number of distinct observations and (ii) the desire to resolve to some extent the short time dynamics.

In order to be able to work with a limited number of distinct observations (i) similar to the other applications in the present work, we consider translationally invariant random unitaries, which also obey lattice inversion symmetry. This restricts our choice to unitaries corresponding to Hamiltonians of the form
\begin{align}
    \hat H=a\big(\hat S_1^+ \hat S_2^-+ \hat S_1^- \hat S_2^+\big)+b\hat\sigma_1^z\hat \sigma_2^z+c\big(\hat\sigma_1^z+\hat\sigma_2^z\big)\ ,
\end{align}
which is split into three blocks corresponding to the possible values of total magnetization of the two spins involved. For the zero magnetization block we sample random eigenvalues $\theta_1,\theta_2$ uniformly from $[-\pi,\pi]$, which translate into couplings $a=(\theta_2-\theta_1)/2$ and $b=(\theta_1+\theta_2)/2$. The diagonal components in the polarized sectors are $b\pm2c$ and we sample uniformly distributed $c$ from $[-\pi/2,\pi/2]$ as the third independent random variable.

Finally, to achieve (ii), we construct the random unitary as $\hat U(a,b,c)=e^{-i\Delta t\hat H}$ with $\Delta t=0.1$. Thereby, all unitaries have a large admixture of the identity and do not fully scramble the two spin subspace within one step. In Fig.~3 the effect of this restriction is that for large numbers of latent variables the maximal test error is only reached after a few steps and not immediately after the first step. We expect that considering larger support of observables would have a similar effect, because then locality of the random unitaries could be resolved.

\begin{figure}[b]
    \centering
    \includegraphics[width=\columnwidth]{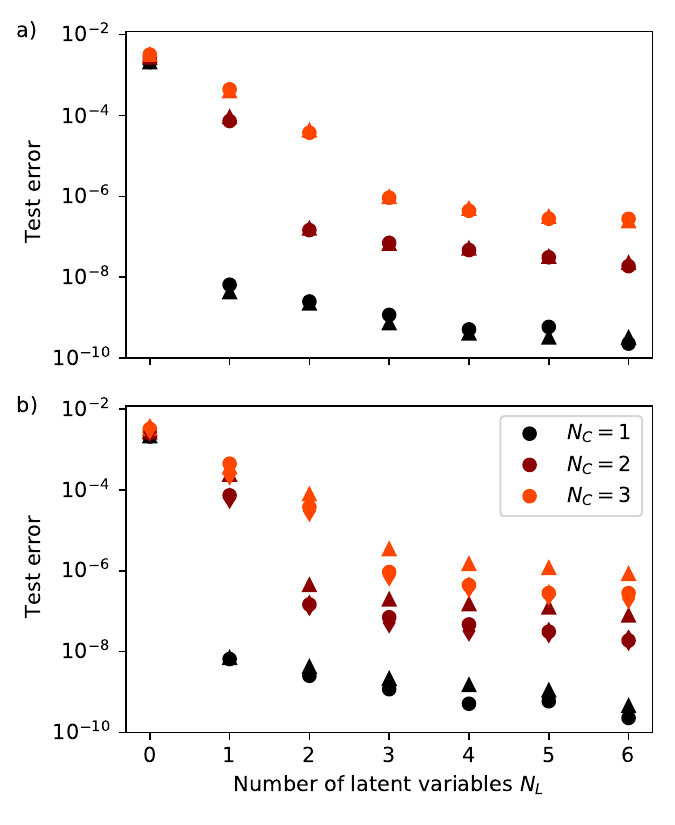}
    \caption{Dependence of the training outcome on varying network size and number of training data. The physical setting is the same as in Fig.~1 of the main text. a) Comparison of network widths 200 (triangles) and 400 (dots). b) Comparison of training data sizes 500 (triangles), 2000 (dots), and 8000 (diamonds).}
    \label{fig:finite_net_and_data}
\end{figure}

\section{Training of autoencoder networks}
All results were obtained using basic feed-forward neural networks exhibiting the bottleneck structure discussed in the main text. A feed-forward neural network made up of layers corresponding to a sequence of alternating affine-linear and non-linear transformations. To obtain the activations $a_j^{(l)}$ of the $l$-th layer, the activatons of the previous layer, $a_j^{(l-1)}$, are transformed as
\begin{align}
    a_j^{(l)}=\sigma\Big(\sum_{k}W_{jk}a_k^{(l-1)}+b_j\Big)\ .
\end{align}
Here, $W_{jk}$ and $b_j$ are variational parameters and $\sigma$ is a fixed non-linear activation function.

Throughout this work we used $\sigma\equiv\tanh$ and networks with two encoder layers between input and latent space, as well as two decoder layers between latent space and output, see Fig. 1 of the main text. Each of these layers had a size of 400 neurons; convergence in network size is demonstrated exemplarily in Fig.~\ref{fig:finite_net_and_data}a. To train the network we used the Adam optimizer \cite{kingma14} with learning rate $r=5\times10^{-4}$ and the standard parameters $\beta_1=0.9$, $\beta_2=0.999$, $\epsilon=10^{-8}$. We chose training batch sizes and the total number of training steps depending on the size of the training data set and convergence behavior. These values are summarized in Tab.~\ref{tab:training_values}. The numbers of data given in this table are the total number of data used, of which 80\% served as training data and 20\% were used as test data. In Fig.~\ref{fig:finite_net_and_data}b we show the effect of varying the size of the training data set exemplarily for the setting of Fig.~1 of the main text.

\begin{table}[b]
    \centering
    \begin{tabular}{c|c|c|c}
         Setting & Num. data & Batch size & Num. steps \\\hline
         (G)GE & 2000 & 128 & 250000 \\
         Lindblad evol. & $\sim200$ & 16 & 250000 \\
         Rnd. unitary & $\sim1000$ & 64 & 50000 \\
    \end{tabular}
    \caption{Training parameters for the different applications. Numbers of training data for Lindblad evolution and random unitary evolution varied slightly for the different physical parameter sets due to the compute time required for data generation.}
    \label{tab:training_values}
\end{table}

To avoid overfitting we adopted an ``early stopping'' procedure, in that we used the parameter sets that resulted in the minimal test error during optimization for our further analysis. While in many cases we observed the typical minimum in the test error at intermediate optimization steps (indicating later overfitting), we found that in cases where few variables fully describe the training data that
perfect generalization is achieved and the test error is always reduced in the same way as the training error.

When comparing different sizes of the latent space, we include results obtained with zero latent variables as a reference. In that case the encoder and decoder are completely decoupled, meaning that the decoder receives no information about the input and the network can only learn the best constant to approximate the data, namely the mean.

\section{Intrinsic dimension}

Here we compare the autoencoder's dimensional reduction to an alternative way of computing the intrinsic dimension $I_d$ of data \cite{wold87,borg05,balasubramanian02,van08,mcinnes18,levina04,campadelli15,facco17,mendes20,lopez18}, using a nearest neighbor method introduced in Ref.~\cite{facco17}, where one calculates the distribution $f(\mu)$ of the ratio of distances to the next-nearest and nearest neighbour, $\mu=r_2(\boldsymbol{x})/r_1(\boldsymbol{x})$, of the vector $\boldsymbol{x}$. In our case, the vector $\boldsymbol{x}$ contains as entries (i) expectation values of all $N_O$ different measured operators or (ii) values of $N_L$ different latent variables.  For data points that are locally uniformly distributed on an $I_d$ dimensional hypersphere, $f(\mu)=I_d \mu^{-I_d-1}$. We can determine $I_d$ by fitting $f(\mu)$ to the input data for the encoder. 

Following Ref.~\cite{facco17}, the explicit algorithm for determining the intrinsic dimension $I_d$ consists of the next steps:
\begin{enumerate}
\item For each vector $\boldsymbol{x}_j$, $j\in[1,N_{real}]$, containing as entries (i) expectation values of measured operators or (ii) values of latent neurons for $j$-th data element, find the Euclidean distance to the nearest and the next-nearest neighbour, $r_1(\boldsymbol{x}_j)$ and $r_2(\boldsymbol{x}_j)$, respectively.
\item Compute the ratio $\mu=\frac{r_2(\boldsymbol{x}_j)}{r_1(\boldsymbol{x}_j)}$ for each of $N_{real}$ different data elements.
\item Get the distribution of $\mu$, $f(\mu)$. For points that are drawn from an $I_d$ dimensional hypersphere with constant density, 
\begin{equation}\label{eq:f_mu}
f(\mu)=I_d \, \mu^{-I_d-1}    
\end{equation}
For the algorithm to work, in our data, uniform distribution must be satisfied just locally, roughly on the level of nearest neighbors.
\item Calculate the cumulative distribution $P(\mu)=\sum_{\mu'<\mu} f(\mu')$
and fit $I_d$ from the linear dependence
\begin{equation}\label{eq::Id_fit}
\ln(1-P(\mu))=-I_d \ln(\mu).
\end{equation}
\end{enumerate}

\begin{figure}
\includegraphics[width=0.9\columnwidth]{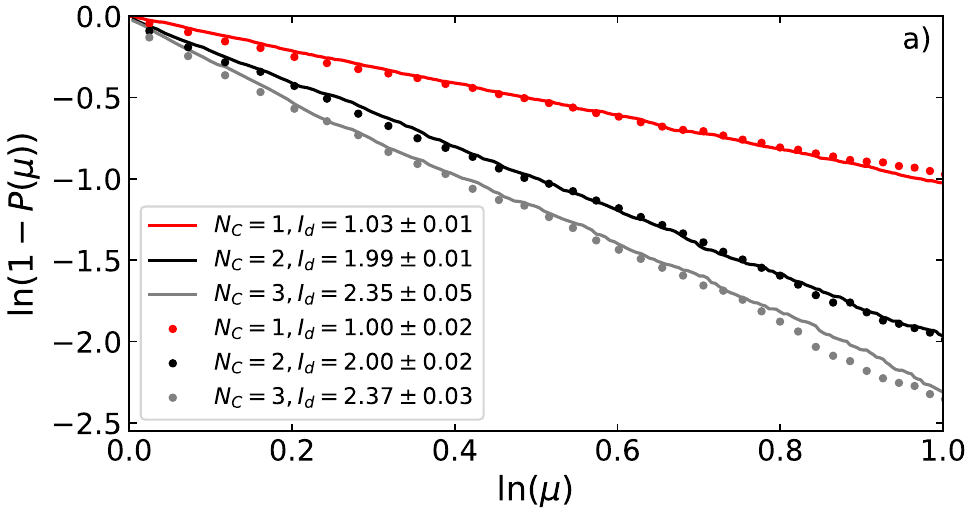}
\includegraphics[width=0.9\columnwidth]{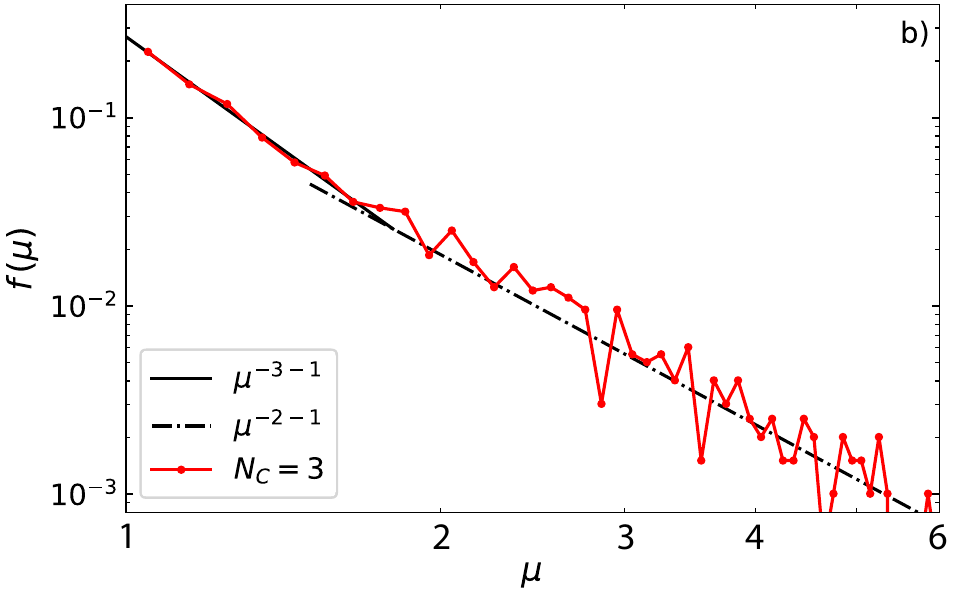}
\caption{
a) Dependence \eqref{eq::Id_fit} for data set with QIM GGE expectation values that are parametrized with $N_C=1,2,3$ Lagrange multipliers, implying $I_d=N_C$. $I_d$ is fitted from the slope. Solid lines correspond to the analysis done directly on expectation values of operators (autoencoder input data) using $N_{real}=2000$ different Lagrange multipliers. Points are obtained using data in the latent space, with a prior dimensional reduction performed by the encoder and $N_{real}=1000$. 
b) Distribution \eqref{eq:f_mu} for $N_C=3$ shows two distinct powers, corresponding to $I_d\approx 3$ at shorter scales ($\mu<1.5$) and $I_d\approx 2$ at longer scales ($\mu>2$).
}
\label{fig:Id}
\end{figure}

Fig.~\ref{fig:Id}a shows dependence \eqref{eq::Id_fit} for sets of expectation values, calculated from GGEs, Eq.~\eqref{eq:sm:GGE}, with $N_C=1,2,3$ charges of QIM ($\hat C_i, i=0,1,2$, Eq.~\eqref{eq:sm:charges}) and random corresponding Lagrange multipliers. Here, we re-analyse the exact diagonalization results for $N=12$ system size and $h_x/J=0.6$, in the main text presented in Fig.1. The above procedure, where $I_d$ is fitted from the linear slope in relation \eqref{eq::Id_fit}, is performed: (i) Directly on the GGE expectation values, i.e., the input data of encoder, for $N_{real}=2000$ different choices of Lagrange parameters (solid lines). Each data element is presented as a $N_{\boldsymbol{O}}$ dimensional vector $\boldsymbol{x}_j$ with entries corresponding to measurements of operators $O(\boldsymbol\alpha)$. Or (ii) on the data from the latent representation with $N_L=4$ and $N_{real}=1000$ training realizations. In this case, each data element is presented as a $N_L$ dimensional vector $\boldsymbol{x}_j$ with entries corresponding to latent values. In the latter case, a prior dimension reduction is performed by the encoder.

Since the data is fully characterized by Lagrange parameters, we expect $I_d=N_C$, which is confirmed for $N_C=1,2$, Fig.~\ref{fig:Id}a. Using Eq.~\eqref{eq::Id_fit} for $N_C=3$, we get $I_d\approx 2.4<3$. The reason for this discrepancy is revealed when we look directly at the distribution $f(\mu)$, Eq.~\eqref{eq:f_mu}. Fig.~\ref{fig:Id}b shows that for $N_C=3$, $f(\mu)$ has two different slopes corresponding to $I_d\approx 3$ at shorter scales $\mu<1.5$ and $I_d\approx 2$ at larger $\mu>2$. In this case, Eq.~\eqref{eq::Id_fit} is not valid and should not be used. We do not have a physical interpretation of our observation, which nonetheless signals that method of Ref.~\cite{facco17} should be used with some caution when data has some extra structure. In these cases, analysing $f(\mu)$ is more appropriate compared to the cumulative distribution $P(\mu)$.

We conclude that $I_d$ can give a good cross-check of the dimensional reduction by autoencoders; however, this approach can be subtle and it is restricted to detecting the effective dimensionality of the data without further insights into the characteristics of the encoded information.

\section{Hamiltonian reconstruction}

\begin{figure}[h!]
\includegraphics[width=0.8\columnwidth]{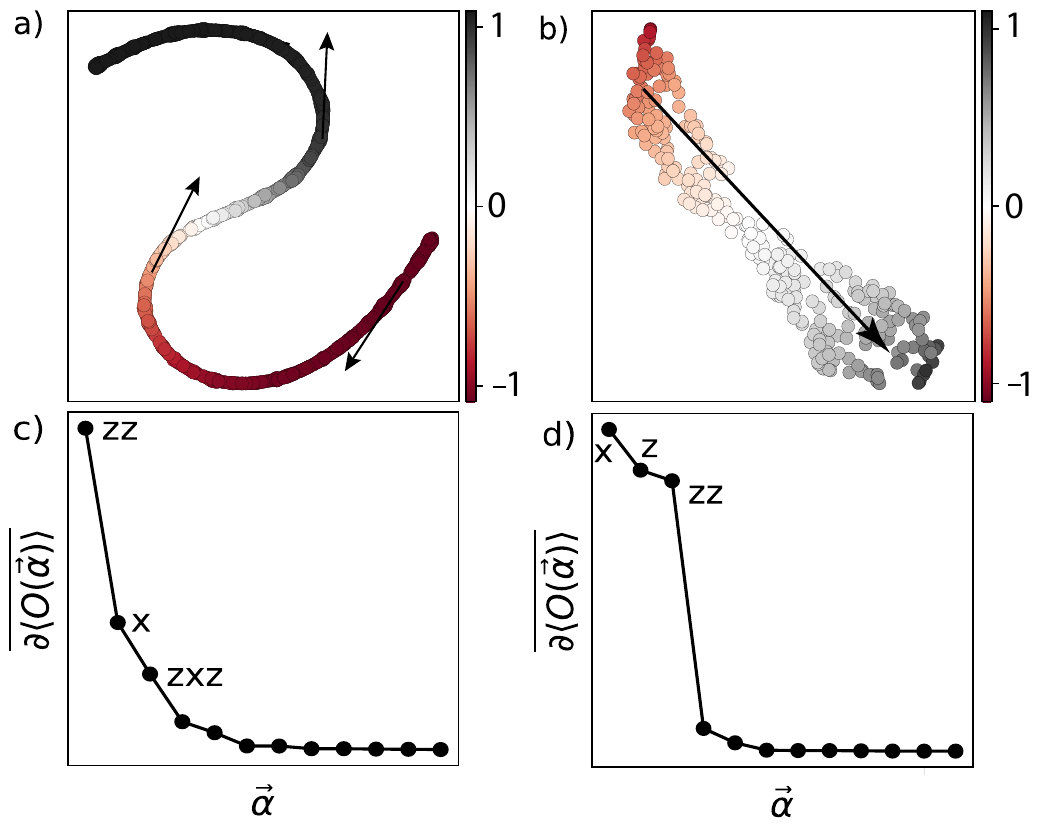}

\vspace{-.2cm}
\caption{
a) t-SNE for exactly thermal states of integrable QIM, Eq.~\eqref{eq::SM-TFIM}. Gradients of measured expectation values $O_\rho(\boldsymbol{\alpha})$ are calculated along the tangential direction and then averaged along the curve. b) t-SNE for nearly thermal states of non-integrable QIM, $\hat H=\sum_{j} J \hat\sigma_j^z  \hat\sigma_{j+1}^z + h_x \hat\sigma^x_j + h_z \hat\sigma^z_j$, obtained at a weak coupling to baths $\epsilon=0.001$. Gradients of $O_\rho(\boldsymbol{\alpha})$ are calculated along the principal direction of PCA, shown with an arrow. c) and d) Averaged gradients $\overline{\partial \ave{O(\boldsymbol{\alpha})}}$ for different operators labelled with $\boldsymbol{\alpha}$ for a) and b) case, respectively. Terms from the Hamiltonian show that largest gradients, in particular if evaluated with respect to high temperature steady states as in b).
}
\label{fig::SM_Ham_recon}
\end{figure}

In this section we give more details on the algorithm for the reconstruction of translationally invariant Hamiltonians, which can be used under the condition that expectation values originate from (approximately) thermal states. A minimal knowledge on whether the energy is (at least approximately) conserved is thus required. The algorithm has the following steps:
\begin{enumerate}
\item Check that a single parameter (related to temperature or $\ave{\hat H}$ by an invertible function) to a good approximation describes all expectation values, i.e., the latent representation must be (roughly) one dimensional. This can be done by calculating the test error as a function of number of latent variables or the intrinsic dimension $I_d$.
\item Calculate the t-SNE of the latent space of a trained autoencoder.

\item Find operators 
$\hat O(\boldsymbol\alpha)\equiv \hat O_{j=1}(\boldsymbol\alpha)=\hat\sigma_1^{\alpha_1}\ldots\hat\sigma_{|\mathcal S|}^{\alpha_{|\mathcal S|}}$ with the largest gradient in measured expectation values $O_\rho(\boldsymbol{\alpha}) = \tr[\hat O(\boldsymbol{\alpha})\hat\rho]$ with respect to their position in the t-SNE.
Operators are labelled by $\boldsymbol\alpha=(\alpha_1,\ldots,\alpha_{|\mathcal S|})\in\{0,x,y,z\}^{|\mathcal S|}$ and $\hat\sigma_j^{0,x,y,z}$ denoting the identity and the Pauli matrices, respectively.
Gradient in $O_\rho(\boldsymbol{\alpha})$ is calculated in two ways: (a) For exactly thermal states, along the local tangential direction of the one-dimensional  t-SNE, see Fig.\ref{fig::SM_Ham_recon}a. (b) For approximately thermal states, along the first component according to the principal component analysis (PCA) of the t-SNE that is not strictly one-dimensional, see Fig.\ref{fig::SM_Ham_recon}b. 

\item To get a scalar, average the gradient values over all points in the latent space. Operators with largest average gradients $\overline{\partial \ave{O(\boldsymbol{\alpha})}}$ are candidate terms in the Hamiltonian
$H=\sum_{j,\boldsymbol{\alpha}} a_{\boldsymbol{\alpha}} \hat O_j(\boldsymbol{\alpha})$.

\item In order to find the prefactors $a_{\boldsymbol{\alpha}}$ and eliminate operators that are actually not in the Hamiltonian, run Newton's method (or some other method for calculation of roots) in order to find zeros of
\begin{equation}
\frac{\tr[\hat O(\boldsymbol{\alpha}) e^{\lambda_0\sum_{j,\boldsymbol{\alpha}} a_{\boldsymbol{\alpha}} \hat O_j(\boldsymbol{\alpha})}]}{\tr[e^{\lambda_0\sum_{j,\boldsymbol{\alpha}} a_{\boldsymbol{\alpha}} \hat O_j(\boldsymbol{\alpha})}]} -  O_\rho(\boldsymbol{\alpha})=0, 
\, \forall \boldsymbol{\alpha}.
\end{equation}
Here, $O_\rho(\boldsymbol{\alpha})$ are the measured expectation values. Thermal expectation values can be efficiently calculated using imaginary time evolution with tensor networks.
We set $\lambda_0\equiv 1$; this implies that we can determine only the relative values of coefficients in $\hat H$. To get the absolute values, we would need additional information, for example, from the dynamics under such a Hamiltonian.

An alternative way to fix $a_{\boldsymbol{\alpha}}$ would be to calculate the kernel of correlation matrix proposed in Ref.~\cite{bairey19}. With a prior knowledge of a few Hamiltonian candidate terms, kernel calculation is very efficient because the correlation matrix is very small.

\item For approximately thermal states, average $a_{\boldsymbol{\alpha}}$ over different realizations to get a better estimate.
\end{enumerate} 

Now, we analyse the data presented in the main text (Fig.1c and Fig.2b) and previous sections. Fig.~\ref{fig::SM_Ham_recon} shows (a,b) t-SNEs  and  (c,d) relative strengths of averaged gradients $\overline{\partial \ave{O(\boldsymbol{\alpha})}}$ for data obtained from Gibbs ensembles ($N_C=1$) of integrable QIM (left) and from steady states of weakly open ($\epsilon=0.001$) chaotic QIM (right).
In the case of exactly thermal states with, $\hat H=\sum_{j} J \hat\sigma_j^z  \hat\sigma_{j+1}^z + h_x \hat\sigma^x_j$, operators $O(\boldsymbol\alpha)$ with $\boldsymbol{\alpha}=zz,x$ have largest $\overline{\partial \ave{O(\boldsymbol{\alpha})}}$, Fig.~\ref{fig::SM_Ham_recon}c. This confirms that our method detects the Hamiltonian terms. However, we also get significant contribution from operators that are products of terms in $\hat H$, i.e., $\boldsymbol{\alpha}=zxz$ and these are eliminated in the step 5. 
For weakly open case with $\hat H=\sum_{j} J \hat\sigma_j^z  \hat\sigma_{j+1}^z + h_x \hat\sigma^x_j + h_z \hat\sigma^z_j$ and pretty high-temperature steady states, only the Hamiltonian terms $\boldsymbol{\alpha}=x,z,zz$ have significant values, Fig.~\ref{fig::SM_Ham_recon}d. 

For exactly thermal states, the algorithm reconstructs the $\hat H$ up to very high precision, basically given by the quality of the Newton's method implementation. Also for nearly thermal expectation values, obtained with a chaotic $\hat H$ and a weak $\epsilon=0.001$ coupling to Lindblad baths, the relative strength of fields is reconstructed up to 5\% and 1\% for $h_x/J$ and $h_z/J$, respectively. This shows that Hamiltonian reconstruction would be possible also for results  from slightly noisy quantum devices (NISQ) or from a condensed matter setup with inevitable coupling to phonons. At larger $\epsilon$, errors grow since steady states are further away from thermal states.
\bibliographystyle{apsrev4-1}
\bibliography{references}

%=================================